# Magnetic Fields in Molecular Clouds—Observation and Interpretation

Hua-Bai Li

Department of Physics, The Chinese University of Hong Kong, Hong Kong SAR, China; hbli@cuhk.edu.hk

**Abstract:** The Zeeman effect and dust grain alignment are two major methods for probing magnetic fields (B-fields) in molecular clouds, largely motivated by the study of star formation, as the B-field may regulate gravitational contraction and channel turbulence velocity. This review summarizes our observations of B-fields over the past decade, along with our interpretation. Galactic B-fields anchor molecular clouds down to cloud cores with scales around 0.1 pc and densities of $10^{4-5}$ $H_2$/cc. Within the cores, turbulence can be slightly super-Alfvénic, while the bulk volumes of parental clouds are sub-Alfvénic. The consequences of these largely ordered cloud B-fields on fragmentation and star formation are observed. The above paradigm is very different from the generally accepted theory during the first decade of the century, when cloud turbulence was assumed to be highly super-Alfvénic. Thus, turbulence anisotropy and turbulence-induced ambipolar diffusion are also revisited.

**Keywords:** star formation; magnetic field; turbulence

## 1. Introduction

In one of the most popular review articles of star formation theories before the 1990s, 'Star formation in molecular clouds—observation and theory' [1], we were told that magnetic fields are the main force regulating the self-gravity that pulls the gas together in a molecular cloud to form stars. In the next review article, 'Theory of star formation', published in the same journal 20 years later [2], it was suggested that the favoured scenario of star formation has moved from magnetic-field regulation to turbulence regulation. At present, slightly less than half-way to the next 20-year review, I believe that the paradigm has changed again: The simple picture that gravitational contraction regulated by one single dominant force fails to explain the new observations of molecular clouds summarized in this article. The observations are focused on the interactions between the B-field and gravity and between the B-field and turbulence.

The former depends on Zeeman measurements (to estimate the field strength) and field-aligned dust grains (to probe field orientations). The Zeeman measurements have been compared with the "textbook" model (Figure 1) after Bayesian analysis [3]. In Section 2, however, I illustrate that both the logic behind the comparison and the Bayesian analysis are problematic. More observations with lower uncertainties and a more sophisticated model are needed. On the other hand, in Section 3, the effects of B-field orientations on cloud fragmentation and star formation efficiencies are clearly shown.

The ordered field orientation should also leave footprints on turbulence velocity fields which, however, have not been observed in the principal component analysis [4]. In Section 4, I explain why and show that an improved method can, indeed, reveal turbulence anisotropy. While the anisotropy is a result of turbulence-field coupling, we observed that the decoupling is significant at sub-pc scales. This may provide a solution for the so-called "magnetic braking" catastrophe for protostellar disk formation, and indicates that ideal MHD may not be ideal for simulations at cloud-core scales.

## 2. Zeeman Measurements

How the B-field strength, $B$, varies with the gas density, $\rho$, has been assumed as a way to understand how dynamically important the B-field is during cloud contraction, based on the logic in Figure 1 (upper-left), which predicts a power law, $B \propto \rho^n$, with a constant index depending on the dynamic importance of the B-field compared to self-gravity. Crutcher et al. [3] have pioneered putting the aforementioned idea into practice using Zeeman measurements. However, the notoriously difficult Zeeman measurements overshadowed the logic loopholes in the plan, which are detailed in Section 2.1. Moreover, many MHD simulations aiming to interpret the Bayesian analysis of the Zeeman measurements have only focused on the index $n$, but failed to reproduce other observed parameters (as pointed out in Section 2.2). This implies that either the Bayesian analysis or the simulations are problematic. We, thus, revisited the Bayesian analysis [5] and, indeed, found that the current data quality cannot pin down $n$, along with other parameters, as summarized in Section 2.3.

### 2.1. Usually Ignored Facts about the B–ρ Relation

**Magnetic criticality and the $B$–$\rho$ relation**

Virial equilibrium, based on the hydrostatic equilibrium of a self-gravitating spherical volume of ideal gas, is frequently used to calculate the maximum cloud mass that can be supported by thermal pressure against self-gravity, which is called the thermal critical mass (or Jean's mass). Similarly, $\underline{M} = |U|$ is usually used to define the magnetic critical mass, $M_B$, where $\underline{M} \approx B^2 \lambda^3/6 = \Phi^2/6\pi^2\lambda$ is the magnetic energy; $\Phi$ and $\lambda$ are the magnetic flux and radius, respectively. $U = -3GM^2/5\lambda$ is the gravitational potential energy, where $M$ and $G$ are the mass and gravitational constant, respectively. More carefully, numerical models are used to consider more general (flattened) cloud shapes and the contribution from magnetic tension to the balance; however, the term that balances with $U$ can still have the format of $c\Phi^2/\lambda$, with some constant $c$ depending on the shape (e.g., [6]).

The condition with a mass above/below $M_B$ is called magnetically super-/sub-critical. Note that $U/\underline{M} \propto (M/\Phi)^2$, which is a constant assuming flux freezing. This has been widely interpreted to indicate that, if a cloud is super-critical, it will remain so. This popular, yet false, consequence ignores the fact that the spherical assumption is impossible to hold, unless $\underline{M} \ll |U|$ (Figure 1, upper-right). Otherwise, the Lorentz force will guide an anisotropic contraction to flatten the gas cloud and the critical $(M/\Phi)^2 \propto c$ varies with time (shape).

A little more consideration of the magnetic tension (~$B^2/R$; where $R$ is the field curvature radius) and pressure (~$B^2/\lambda$) forces [7] indicates that the forever-super-critical picture is naïve. Both magnetic tension and pressure are inversely proportional to the scale of the fifth power during contraction (note that $B$ is inversely proportional to the cross-sectional area $\propto \lambda^2$), while gravitational force is inversely proportional to the second power. Thus, the contraction will slow down in the direction anti-parallel with the magnetic tension/pressure force, which is perpendicular to the mean field. This means that the index $n$ decreases with time, as illustrated by the MHD simulations in Figure 1 (upper-right) [7]. Therefore, the assumption of a constant $n$ is unrealistic.

**Spatial and temporal $B$–$\rho$ relation**

Note that the $B$–$\rho$ relation, discussed above, is for one cloud core along the time axis. Let us call this the "temporal $B$–$\rho$ relation", which is not directly observable. The Zeeman $B$–$\rho$ relation (Figure 1, lower-left) is based on a collection of clouds/cores at various positions at the same time. Let us call this the "spatial $B$–$\rho$ relation". Another type of observable spatial *B–ρ* relation is based on various density structures within one cloud (e.g., Figure 1, middle-right); again, at the same time.

*2.2. Simulations Aiming to Reproduce the Zeeman B–ρ Relation*

Recent simulations of self-gravitating, magnetized, and turbulent clouds (e.g., [8–12]) have explored the *spatial B–ρ* relation.

Mocz et al. [11] found the index *n* to be 2/3 for Alfvén Mach number $M_A \gtrsim 1$ and ½ for $M_A < 1$. Li et al. [10] found a 2/3 index in the case of a moderately strong initial magnetic field ($M_A = 1$), and marginally so for a weak initial field ($M_A = 10$). We [12] found that the index can achieve 2/3 when the density is above $10^{4.5}$/cc, even with $M_A = 0.6$ (Figure 1, middle-left); that is, the 2/3 index is not necessarily a signature of a "weak" field, compared to the turbulence. Additionally, all of the above simulations require a density well above $10^4$/cc to achieve an index of 2/3, regardless of $M_A$; however, the critical density estimated by the Bayesian analysis [3] was merely 300/cc. This inspired us to revisit the Bayesian analysis, as summarized in Section 2.3.

Most importantly, the 2/3 index from the aforementioned simulations is not even comparable to the textbook model (Figure 1, upper-left). In the upper-right of Figure 1, we put the textbook model into numerical simulations, which illustrated the discussion in Section 2.1 very well: The index n is not a constant of time, it decreases as the field-line density and curvature increase during cloud contraction, unless the mass is more than an order of magnitude larger than the critical mass, $M_B$.

The critical *M/Φ* ratio discussed in Section 2.1 can be reduced to *N/B*—the column density to field strength ratio— for the column density projected along the field. The *B-N* relation is observable as the upper envelope in Figure 1, lower-right panel. The gas accumulated along the field can move a sub-critical cloud towards the right horizontally in the *B–N* plot, until the cloud becomes critical. The slanted part of the *B–N* relation starts right at the critical column density of the flat-top *B* (Figure 1, lower-right) and the upper envelope of the distribution can be very well fitted by the critical *N*. Were the observed cores highly super-critical, the upper envelope would not be this closely aligned with the theoretical critical value; in other words, the observed structures range from magnetically sub-critical (flat-top part in the *B–N* relation) to trans-critical (slanted part in the *B–N* relation).

Thus, the observed or simulated 2/3 index of the *spatial B–ρ* relation is different from the textbook *temporal* 2/3 index from an isolated, highly super-critical cloud (Figure 1; upper-right). Then, how is the spatial 2/3 index achieved for trans-critical structures? I believe that it is due to the fact that real cloud structures are not isolated, unlike in the textbook model. When simulating isolated clouds, trans-critical clouds start with an index close to 2/3, which is later flattened as the field lines are bent by gravity, which increases $M_B$ (Figure 1, upper-right). Gas accretion along magnetic fields of a non-isolated cloud structure allows the structure mass to keep up with the increasing $M_B$, and, thus, it can maintain the index without being highly super-critical. We are carrying out numerical simulations to test this proposal.

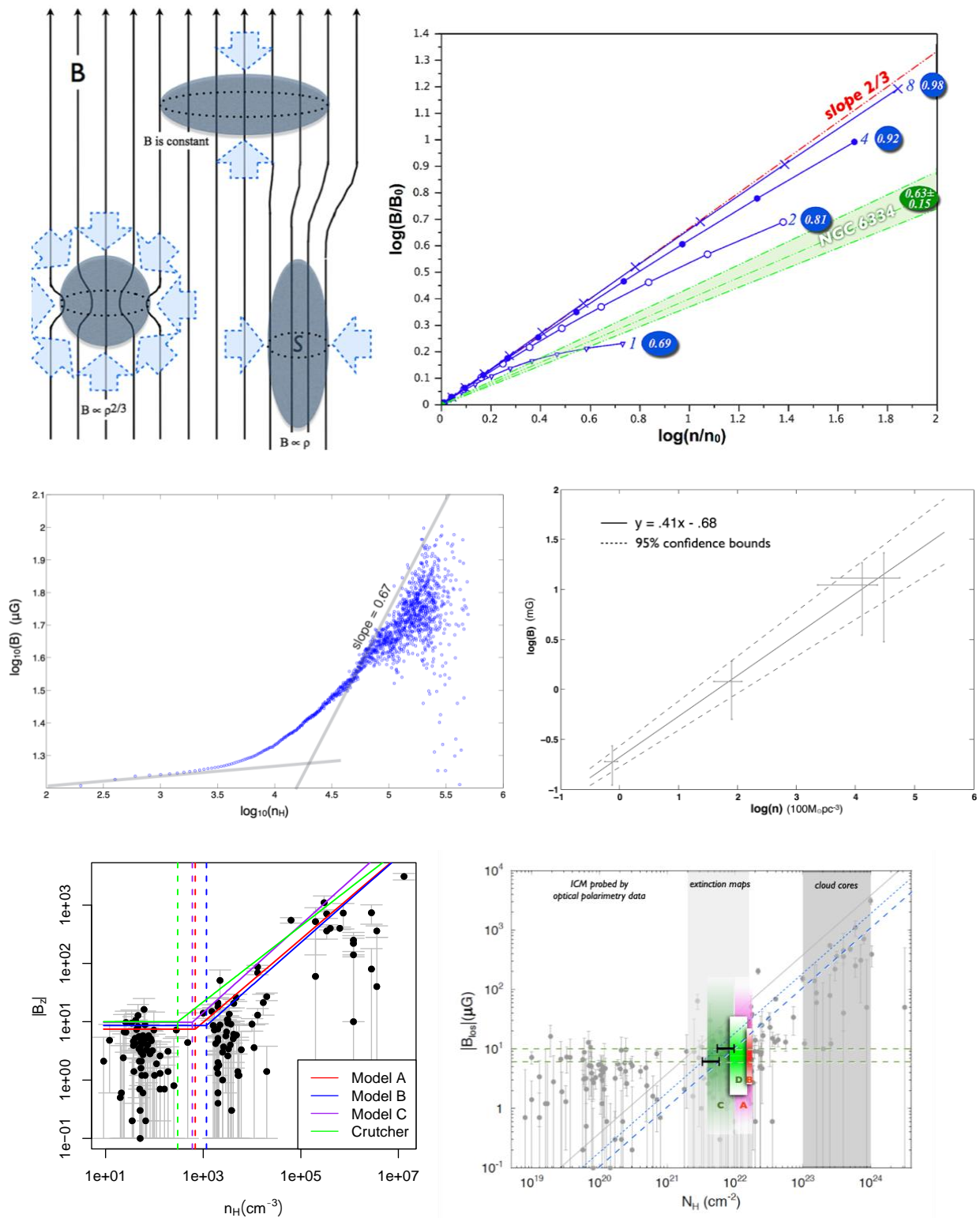

**Figure 1.** Models, simulations, and observations of cloud $B$–$\rho$ relations. (**Upper-left**): The "textbook" model (e.g., [13,14]) of three extreme *temporal B–ρ relation* cases: compression of fixed cloud mass isotropically (**left**) or perpendicular (right)/along (middle) B-fields. Each case results in a particular $B$–$\rho$ relation. Pure perpendicular compression should not happen for gravitational contraction. Contraction isotropically or along B-fields happens for extremely weak or strong fields, respectively; see simulation illustration in the right panel). (**Upper-right**): *Temporal B–ρ relations* from MHD simulations of initially spherical isolated clouds with uniform B-fields. The cloud masses are, respectively, 1, 2, 4, and 8 times the $M_B$ in the initial condition. The numbers in the "ovals" are the minor-to-major axes ratios of the oblate cloud at the end of the simulations [7]. The decreasing in the slopes is due to the fact that, while $M/\Phi$ is a constant (flux freezing),

the "critical *M/Φ*" is increasing as the cloud is flattened (Section 2.1). The NGC6334 (green) data are from observations (middle-right). (**Middle-left**): The *spatial B–ρ* relation from the MHD simulation of a patch inside an isothermal molecular cloud with sub-Alfvénic turbulence [12]. Each data point shows the medium field strength of all the pixels within a density bin. (**Middle-right**): The *spatial B–ρ* relations based on the polarimetry data of NGC 6334 [7]. The density unit used here is hundred solar mass per cubic pc. (**Lower-left**): Fitting various Bayesian models for the Zeeman measurements [5]. Model C is almost identical to Crutcher's [3], but with no limit on the slope index *n* (Crutcher et al. limited *n* to be under 0.75). Model B further releases the restriction on the density uncertainty (factor of 2) used by [3]; see [5] for Model A. Note that all models assume a slope-changing density of $\rho_0$. However, the slopes from simulations (e.g., (**middle-left**)) gradually increase from $n_H \sim 10^3$/cc; there is no single threshold, as assumed in these Bayesian models. (**Lower-right**) Based on the same Zeeman data set used on the left, but with the B-field versus column density [15]. The two horizontal green dashed lines indicate the range of the "flat-top" field strength, fitted in the left panel. The two slanted blue lines indicate the critical column density, given the field strength (dotted) or given the field strength and trans-Alfvénic turbulence velocity (dashed). The region within the two green lines and the two blue lines falls within the column density range (C [16] and D [17]) where the column density probability density functions (PDFs) change from log-normal to power-law. The ranges A [18] and B [19] are the observed thresholds of star formation and are related to the star formation efficiencies in Figure 7.

### 2.3. Bayesian Analysis

The scatter *B–ρ* plot in Figure 1 (lower-left) includes the most Zeeman measurements ever made. Crutcher et al. [3] concluded that a dynamically relevant B-field during core formation is inconsistent, as the upper limit of the *B–ρ* logarithmic plot has a slope (*n*) of 2/3, which, however, is based on problematic logic, as discussed above. Criticisms in the literature have focused on the data itself. For example, Figure 1, lower-left, contains dark clouds and massive cluster-forming clumps; as the former are not likely to develop into the latter, using them to infer an isotropic collapse is questionable. Others are concerned with either B-field morphologies [20] or clump shapes [21].

There had been no attempt to examine the Bayesian analysis until Jiang, Li and Fan [5]. They had both physical and statistical motivations. The authors in [3] concluded that the index of the *B–ρ* relation is zero below some threshold density ($\rho_0$), becoming 2/3 above the threshold. Besides that, the $\rho_0$ derived by [3] is much lower than the value from simulations, as discussed in Section 2.2. The authors in [5] found that some parameters in the Bayesian analysis performed by [3] were overly restricted. The most important is the uncertainty of *ρ*, which was fixed within a factor of two (R = 2) by [3]. R = 2 is certainly too optimistic, based on both theoretical models [22] and observations (e.g., [21]). We [5] kept R as a parameter to fit in our new Bayesian analysis. With numerical simulations, we also studied how frequent the true parameter value underlying the data set was covered by the posterior distribution for various R (more details are given in the next paragraph). The Bayesian information criterion (BIC; [23]), $\ln(s)p - 2\ln(l)$, was used to judge which model is better, where *p* is the number of parameters in a model, *s* is the sample size, and *l* is the maximal likelihood of the model. A smaller BIC value is preferred. Our simple modifications [5] reduced the BIC of the model of [3] by more than 100.

**Errors-in-variables models**

Note that, as both *B* and *ρ* come with significant uncertainty, we are facing the so-called *errors-in-variables* regression models. It is well-known by statisticians that, when the regressors are measured with errors, the parameter estimates do not tend to be the true values, even with very large samples. Arguably, the most important impact of [5] was to remind astronomers that they are facing an errors-in-variables regression when trying to derive the *B–ρ* relation with Zeeman measurements. We [5] illustrated this with simulated data.

In our simulations, R varied between 1 and 40, in order to represent different uncertainty levels for the *ρ* observations. Other parameter values were sampled from the prior distributions assumed in the models of [3]. The simulated observations of LOS fields and number densities were also generated, according to the Bayesian models. If a Bayesian algorithm is effective, the true values underlying the simulated data should be covered by the posterior distribution of the corresponding parameters. The coverage rate of the 95% highest posterior density interval (HPDI) of each parameter in the model should not be significantly lower than 95% if the algorithm is effective [24]. The results are shown in Figure 2, adopted from [5], where *Model C* is almost identical to that of [3]. We [5] released the restriction of R = 2 in *Model C* to create *Model B*. We can see, from Figure 2, that R can be very well-estimated by *Model* B, and so can $B_0$ (the flat-top field strength with a density below $\rho_0$). Model B dramatically improved the estimate of $\rho_0$ from Model C but, still, the HDPI coverage rate was significantly lower than 95%. Finally, and most importantly, it is not possible to estimate *n* from any of the models, unless R < 2. With the above knowledge, which is also the main improvement from [3] to [5], we can properly interpret the results of the Bayesian analysis.

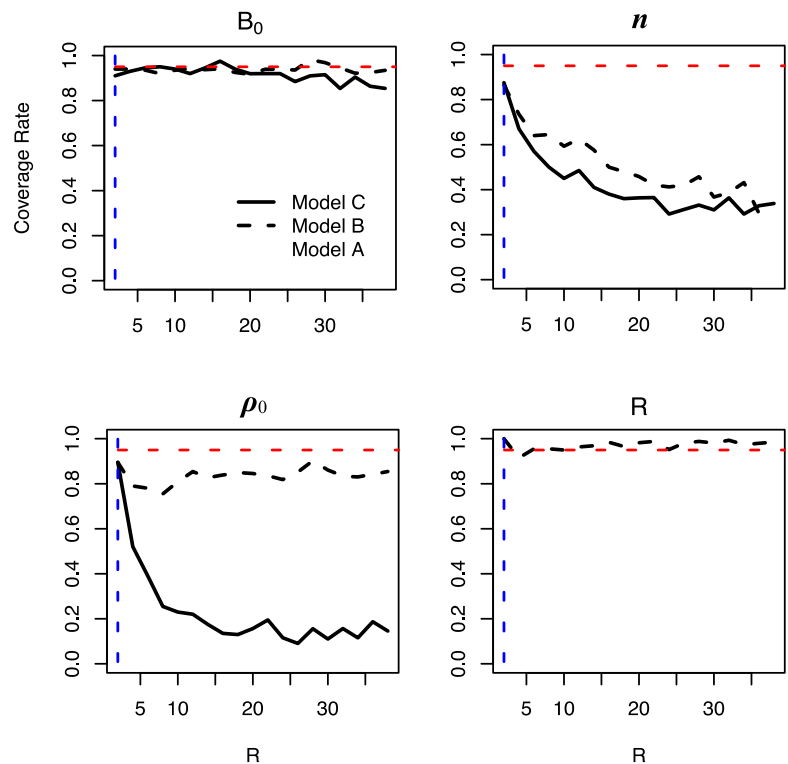

**Figure 2.** Coverage rate of the 95% HPDI of each parameter under different uncertainties (R) of $\rho$ [5]. The red dash line corresponds to the target coverage rate of 95%, while the blue dash line marks R = 2. Model B (see the text) has the best estimates of $B_0$, $\rho_0$, and R; however, the index *n* is not estimable by any model when R > 2.

### *n*, the index of the *B*–$\rho$ power law

Tritsis et al. [21] compared the volume densities adopted by [3] with those appearing in the literature and found differences by factors between 2 and 60 with a mean at 15, if the outlier (Sgr-B2) with R ≈ 400 is excluded. After releasing R from being fixed at 2 in [3] to a parameter in the Bayesian analysis, we [5] estimated R to be 9.3 [2.4, 19.0] with Model B, where […] shows the 95% HPDI. As this estimate of R is reliable (Figure 2) and significantly higher than 2, we should not draw any conclusions about *n* with the current data set. Note that, when R > 2, no model can give a reliable estimate of *n* (see Figure 2)

### $\rho_0$, the transition from magnetically sub- to super-critical

Our [5] estimate of $\rho_0$ (Figure 1, lower-left) was significantly differently from that of Crutcher et al. [3]. Instead of 300/cc [3], *Model* B obtained a peak at 1125/cc within a wide 95% HPDI, [366, 2616], which relieves our concern mentioned in Section 2.2, to some degree—now, $\rho_0$ is one magnitude closer to the MHD simulations. In Figure 1, middle-left, the *B*–$\rho$ relation derived from MHD simulation has no sharp turning point but, instead, has a gradually increasing slope, which takes around two decades of density to grow to 0.67, corresponding to the wide HPDI. This also offers an alternative explanation (besides the unreliable Bayesian estimate) of the "discrepancy" between the *n* values inferred from the Bayesian model (0.67) and from the polarimetry data (0.4; Figure 1, middle-right). If we fit data above 1000/cc in Figure 1, middle-left, which covers the "slope transition" densities from the simulations, both the simulation (middle-left) and observation (middle-right) give a slope ≈ 0.4.

## 3. Grain Alignment Measurements and Implication

Dust grains in molecular clouds cause the 'extinction' of background starlight. The residual background starlight (i.e., the light observable after dust extinction) is polarized. This is because elongated dust grains are aligned by the B-fields, such that the longer axes tend to be perpendicular to the fields [13], which results in a greater degree of extinction for the polarization perpendicular to the fields and leaves the residual light polarized along the fields. The polarization fraction depends on the dust density, shape, and grain alignment mechanisms along the entire line-of-sight between the observer and the background star. Thus, it is important to keep in mind that the method cannot probe a scale smaller than the stellar distance. A clear trend observed is: The further the star, the higher the chance for the polarization to be closely aligned with the galactic disc. This means that, at the scale of a few hundreds of parsecs, the field directions have some structures that significantly deviate from the galactic disc. The effects of these directional structures on star polarization tend to cancel each other when the line-of-sight is much larger than 500 pc [15]. Note that the cloud accumulation length is a few hundred parsecs, such that cloud B-fields, while correlated with the large-scale galactic field [25], should not be perfectly aligned with the galactic disc direction, as even 100-pc scale galactic fields are not perfectly aligned with the galactic disk, not to mention cloud fields. On the other hand, the 100-pc scale galactic fields and cloud fields have been shown to be highly correlated, down to 0.1 pc [26], which requires the turbulence in the cloud bulk volume to be trans- to sub-Alfvénic. We discuss the consequences of this ordered field on cloud fragmentation and star formation in Section 3.2. In Section 3.1, we check whether the field is still ordered below 0.1 pc.

*3.1. Anchoring Galactic B-Field in Cloud Cores?*

Li et al. [26] compared the magnetic field directions inferred from polarimetry data obtained from 100-pc scale inter-cloud media (ICM) and from sub-pc scale molecular cloud cores. The former were inferred from star-light polarization and the latter were based on dust thermal emission, as the dense cores do not allow for the penetration of background stars light but have strong dust thermal emission, which is polarized in the direction along the dust long axes. The highly correlated result led us to conclude that cloud turbulence must be sub-Alfvénic. A similar result has been observed in extragalactic clouds [25]. This scenario was later confirmed by numerical simulations (e.g., [10]) that field alignment seen by Li et al. [26] (Li09, hereafter) is possible only if the turbulence is trans- or sub-Alfvénic. However, recently, when we extended the Li09 study to a smaller scale (<0.1 pc) using interferometer data [27,28], we saw that the B-field significantly deviated from the field directions in the surrounding ICM (Figure 3) [12]. An apparent question to ask is whether this high-resolution result contradicts the sub-Alfvénic picture concluded earlier.

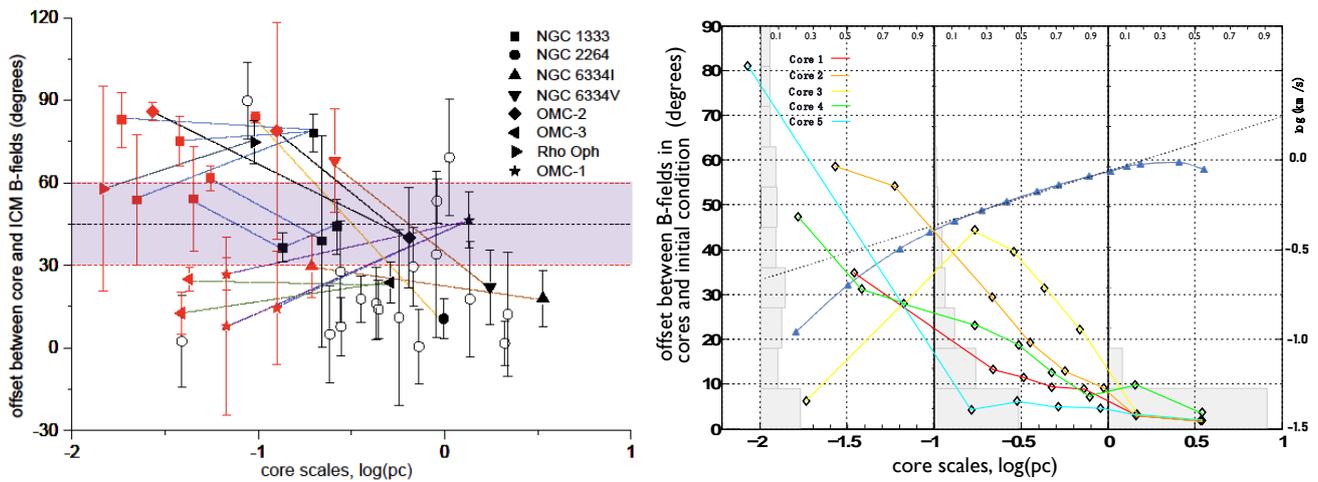

**Figure 3.** (**Left**)*:* Core/ICM B-field directional offsets against the spatial scales of cores [12]. The black symbols are from the data in Li09 and the red symbols are interferometer data [27,28]. Data from Li09 with no interferometer counterparts are shown as hollow circles. The error bars indicate the interquartile ranges (IQRs), not error of the mean, of the B-field orientation distributions. Above ~0.1 pc, most of the offsets are smaller than 45 degrees. The upper envelope grows with decreasing scale, and, below 0.1 pc, there seems to be no polarizational direction preference at all. (**Right**)*:* The results of sub-Alfvénic cloud MHD simulations [12]. The diamonds show the 3D core field directions (measured from the initial uniform field direction; left axis) against the core sizes (bottom axis). Each of these 3D offsets is projected along 135 directions, evenly distributed over a $4\pi$ sr solid angle. The scales are binned into "above 1 pc", "below 0.1 pc", and "in between". The histograms present the fraction distribution of the projected 2D offsets in each bin (the upper axis is the fraction). Note that the 2D histograms are "folded" at 90 degrees (but the 3D offsets are not!), in order to emulate the disability of polarimetry observations in distinguishing supplementary offset angles. Additionally, the turbulent 3D velocity spectrum is plotted (the dark blue line; the velocity is shown in the right axis), which peaks at 2.4 pc (the turbulence driving scale) and is numerically dissipated below 0.1 pc. The dotted line (y = 0.90 $x^{0.31}$) is fit to the velocity spectrum between 0.1 and 1.25 pc.

**The emergence of super-Alfvénic cores from sub-Alfvénic clouds**

We [12] attempted to understand this by simply observing sub-Alfvénic simulations, which can already reproduce the Li09 observation, at scales smaller than 0.1 pc. Indeed, we observed something comparable to the observation (Figure 3): the field directions below 0.1 pc seemed not to be so correlated with the field direction in the initial condition. Note that the observations (left) and simulations (right) in Figure 3 are not directly comparable, as the simulations show the 3D offsets, while the observations are 2D offsets projected onto the sky. To make them more comparable, all the simulated 3D angles were uniformly projected to a $4\pi$ sr solid angle. The histograms of all the projected angles at >1-pc scale, < 0.1-pc scale, and in-between, respectively, are also shown in Figure 3. The distribution from the scales < 0.1 pc is very flat, comparable to the observations.

Note, however, that the flat histogram in Figure 3 stems from a non-random B-field, which is still correlated with the initial condition, as the 3D offsets were all below 90 degrees. Thus, even though the observed 2D offsets (Figure 3, left) almost evenly covered the entire 0–90-degree domain, one cannot conclude a random field from it. Due to the "directionless" property, the polarimetry directional offset cannot exceed 90 degrees; thus, 0–90 and 0–180 (random) degree B-field offsets look the same, through the lens of polarimetry.

Still, we need to understand why the field deviation—though not random—increases with the density. For this purpose, we studied the Alfvén Mach number ($M_A$) at various densities and scales (Figure 4) [12]. While the overall $M_A$ was less than unity (red in Figure 4), $M_A$ was distributed as a power function of density at small scales (sub-pc; blue in Figure 4), which can go beyond unity when $n_H$ > 1000/cc. At the typical density seen by interferometers ($n_H$ > $10^5$/cc), $M_A$ can go beyond 3, which can result in a field deviation as large as ~90 degrees. The B-field channelled gravitational contraction, which can focus the kinetic energy without significantly compressing the B-field, must play a major role in the $M_A$ enhancement.

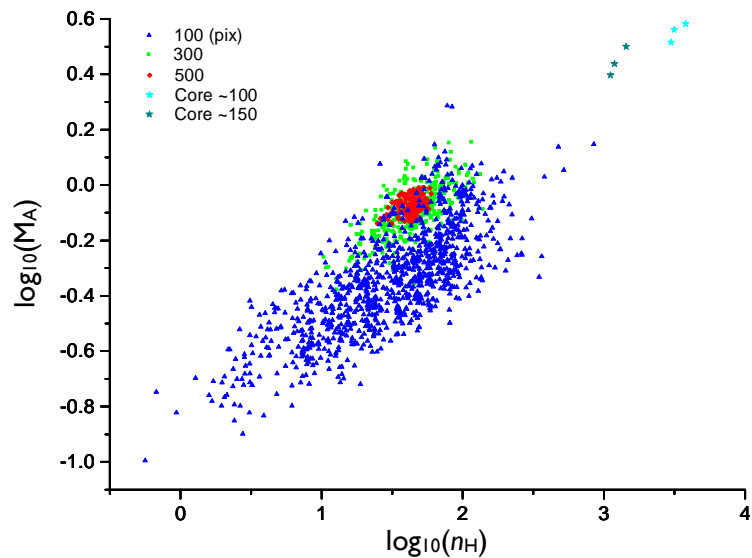

**Figure 4.** Alfvén Mach number versus density at various scales (100-, 300-, and 500-pixel; 500 pixels is equivalent to 2.4 pc in the simulation) from the same MHD cloud simulation in Figure 3 [12]. The simulation cube is fully and evenly sampled by sub-cubes of various sizes, and the $M_A$ and mean density of each sub-cube are plotted. The properties of the three dense cores at 100- and 150-pixel scales are also noted.

**Polarization holes**

The trend of increasing field offsets with decreasing scale (and, thus, with increasing density) explains the so-called "polarization holes"—the tendency of the decreasing fraction of sub-mm polarization with growing column density (*N*). Li09 proposed that polarization holes can occur naturally, due to more B-field structures along lines-of-sight with higher *N*; not necessarily due to the lower grain alignment efficiency in high-density regions, as assumed in the literature. Figures. 3 and 4 support this proposal. The fact that interferometers see a higher polarization fraction than single-dish telescopes [29] also agrees with this proposal. Higher *N* usually implies a line-of-sight going through a higher density and, thus, higher $M_A$ and more B-field directional dispersion can be expected (Figures 3 and 4). When these richer field structures are not resolved, they appear as a lower polarization fraction.

*3.2. The Effects of B-Field Orientations on Cloud Fragmentation and Star Formation*

Rigid field orientations will inevitably leave their touch on the orientations of cloud structures, which have also been recently surveyed, using both optical and sub-mm field orientation tracers (e.g., [15,30]). With starlight polarization data [31], Li et al. [15] studied the alignment between the global cloud orientations (derived from the autocorrelations of extinction maps) and their nearby ICM fields of 13 Gould Belt clouds. They found a bimodal distribution, in which clouds tend to be either parallel with or perpendicular to the mean magnetic field directions of local ICM (Figure 5). Later, the Planck Collaboration [30] found that the relative orientation between the magnetic field and sub-cloud structures at the 10′ scale (limited by the angular resolution of Planck) tends to move away from parallelism with increasing column density (Figure 5). However, these two results may confuse some people; some may even think that the findings contradict each other. Gu and Li [32] revisited the two studies and made a connection between them. It should not be difficult to imagine that the bimodal alignment between the B-fields and the cloud global orientations have impacts on cloud fragmentation [33,34] and, even, star formation efficiencies [35].

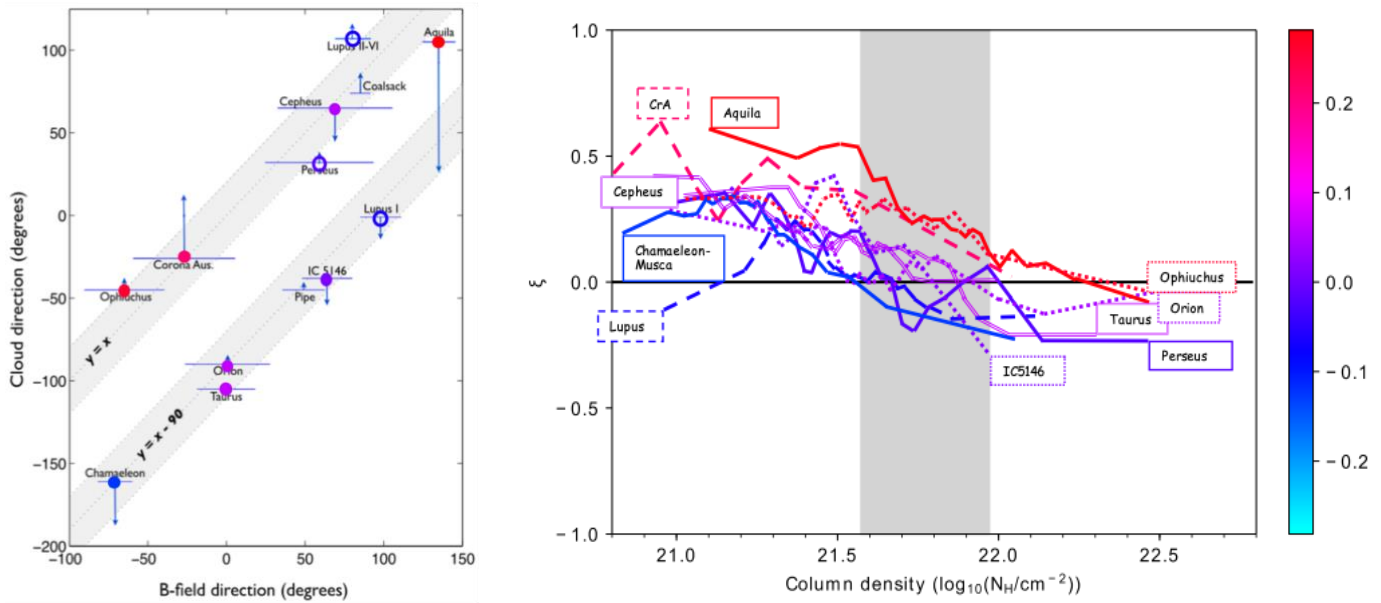

**Figure 5.** (**Left**): The bimodality of the global cloud-field alignment [15]. The horizontal bars show the interquartile ranges of the field directions (not the mean uncertainty), while the arrows show how the cloud orientation changes from low (end) to high (head) density. (**Right**): The pixel-by-pixel field-cloud structure alignment varies between parallel (ξ = 1) and perpendicular (ξ = −1) [32]; ξ is defined by the ratio *(PA − PE)/(PA + PE)*, where *PA* (*PE)* is the population within 22.5° from parallelism (perpendicularity) [30]. The colour code is based on the averaged ξ value of each cloud within the "grey" density range, where the cloud column density PDFs deviate from log-normal and turn towards power-law. The grey range is the same as "zone C" in Figure 1 (lower-right), from Kainulainen et al. [16]. The colour of each cloud is also represented in the left panel. Generally speaking, the perpendicular clouds in the left panel have smaller ξ, which provides the connection between the two different alignment studies. Lupus and Perseus are hollowed, in the left panel, to suggest that they should be excluded from the comparison. Lupus was split into two subregions in [15], but not by [30]. Perseus shows evidence that optical and sub-mm polarimetry data emphasize very different regions [32,35].

**Li et al. [15] vs. Planck collaboration [30]**

Gu and Li [32] first repeated the study of Li et al. [15] using Planck thermal dust polarization data [30] as the B-field tracer, in order to see whether the bimodal alignment remained, and obtained a positive result. This was not a surprise, as the cloud field orientation traced by Planck and optical data are largely consistent (Figure 6). The Pearson's correlation coefficients and p-values (the probability that the correlation is a random result) between the cloud-field offsets based on Planck and starlight data were, respectively, 0.875 and 1.94 × 10$^{-5}$ with the Perseus cloud; and 0.918 and 3.67 × 10$^{-6}$ without Perseus. The reason to exclude Perseus was that the sub-millimetre polarization here is highly affected by stellar feedback [32].

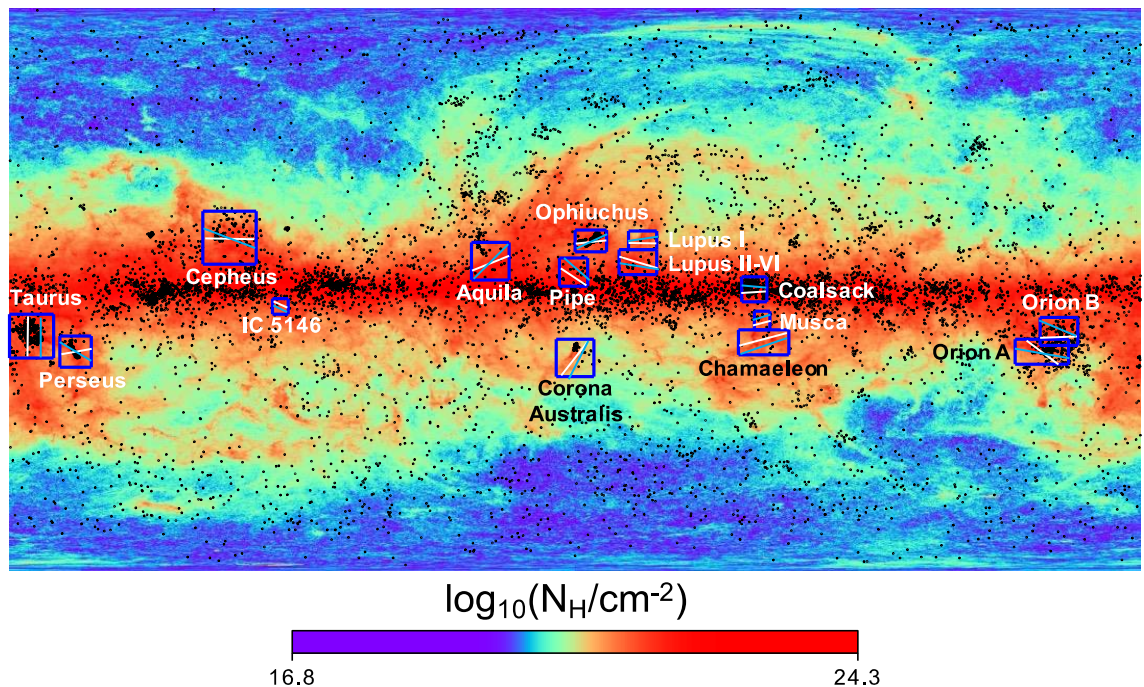

**Figure 6.** Spatial distribution of all stars (black dots) in [31]. The colours represent the total gas column density (in logarithmic scale). The blue rectangles show the locations of all the Gould Belt clouds. Blue and white line segments show the average B-field orientations, inferred from starlight and Planck 353 GHz thermal dust polarization, respectively [32].

After making sure that the field directions inferred from Planck and optical polarimetry were largely the same, Gu and Li [32] attempted to find the connection between the global cloud-field alignment [15] and the Planck pixel-by-pixel cloud structure-field alignment [30]. They studied the pixel-by-pixel alignment in the density range of $A_v$ = 2–5 mag, where the cloud column density PDFs deviate from log-normal and becomes power-law-like [16]. It is generally believed that this density range defines the bounded volume of molecular clouds and, thus, is related to the global cloud shapes studied by [16]. As shown in Figure 5, within $A_v$ = 2–5 mag, the pixel-by-pixel alignment may be closer to either parallel (red) or perpendicular (blue). This means that, at the cloud boundaries, both tendencies are possible. Moreover, this pixel-by-pixel alignment on the cloud borders has some correlation with the global cloud-field alignment, as shown in Figure 5. Those tending to be parallel/perpendicular with the B-field on the borders also tend to globally align in the same way.

**The effect of cloud-field alignment on star formation efficiency**

It would be interesting to check whether the different global cloud-field alignments can affect the cloud fragmentation and even star formation efficiency. The latter is easier to check, as there already exist Gold Belt star formation rate surveys in the literature.

We can expect the effect on star formation, as different cloud-field alignments possess different magnetic fluxes. The same filament will have a higher magnetic flux when oriented perpendicular to the B-field than when aligned with the B-field, due to the different cross-section areas perpendicular to the B-field. Thus, one should expect higher star formation efficiencies from filaments aligned with the B-field, due to the reduced support from the B-field against gravitational contraction. In Figure 7, we compare the star formation rate per unit mass observed by Heiderman et al. [19] and Lada et al. [18] with the cloud-field direction offsets derived with the mean cloud field directions in Figure 6. Indeed, we see the expected correlation between star formation efficiency and cloud-field alignment.

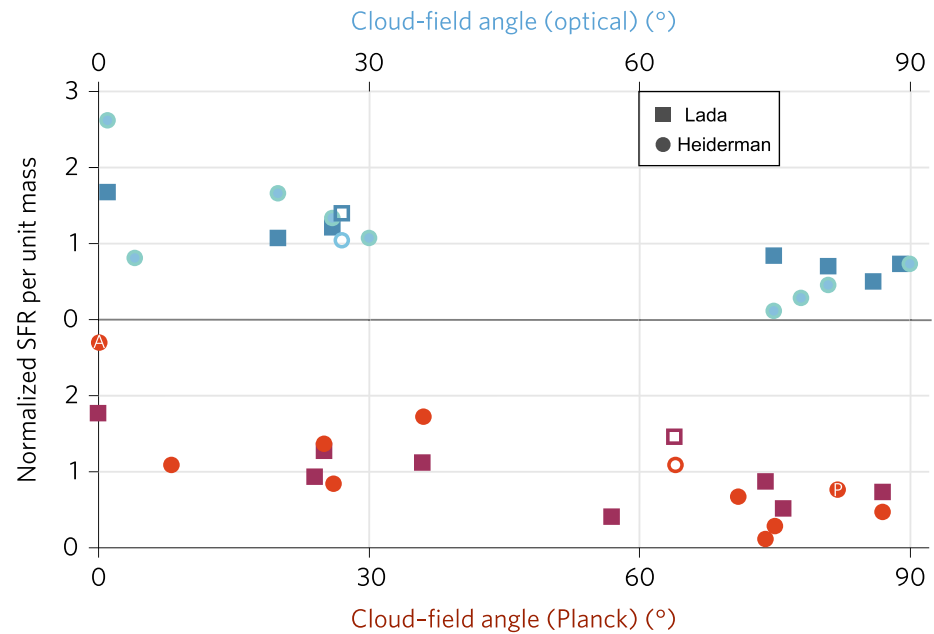

**Figure 7.** Star formation rate (SFR) per unit mass versus cloud-field directional offset for Gould Belt clouds [35]. The SFR are from Heiderman et al. [19] (circles) and Lada et al. [18] (squares). The "cloud-field angle" is the directional difference between the B-field and cloud elongation. The B-field directions are inferred from optical (blue) and sub-mm (red) polarimetry [35], while the cloud elongations are based on the autocorrelation of the extinction data [15]. All four possible combinations of SFR/mass vs. angle show anti-correlation. The SFR from [18] comes without uncertainties, so we [35] performed a permutation test and Spearman rank correlation test of a negative correlation. For the data from [19], which is with uncertainties, the same tests were performed with bootstrap samples from log-normal distributions with the uncertainties as the standard deviations. All test results (four data combinations × two tests) are significant (with very low P-values; see [35] for details).

**The effect of cloud-field alignment on cloud mass cumulative functions**

After discovering the correlation between field orientations and star-formation efficiencies, it is natural to expect the effects of B-fields on fragmentation and, so, we initiated a series of studies [33,34]. Here, we summarize the case of the mass cumulative function (MCF) [34]. A cloud elongated closer in the field direction tends to have a shallower MCF slope; in other words, a higher portion of the gas at high density. The MCF($A_V$) is defined by the ratio of the total mass above extinction $A_V$ to $M_{trs}$, which is the mass above the extinction, $A_{Vtrs}$, where the extinction PDF starts to deviate from log-normal (which is an indication of gravitational bounding). Thus, MCF($A_{Vtrs}$) = 1. Setting MCF($A_{V10\%}$) = 0.1, the MCF slope can be defined within [$A_{Vtrs}$, $A_{V10\%}$] as shown in Figure 8 (right).

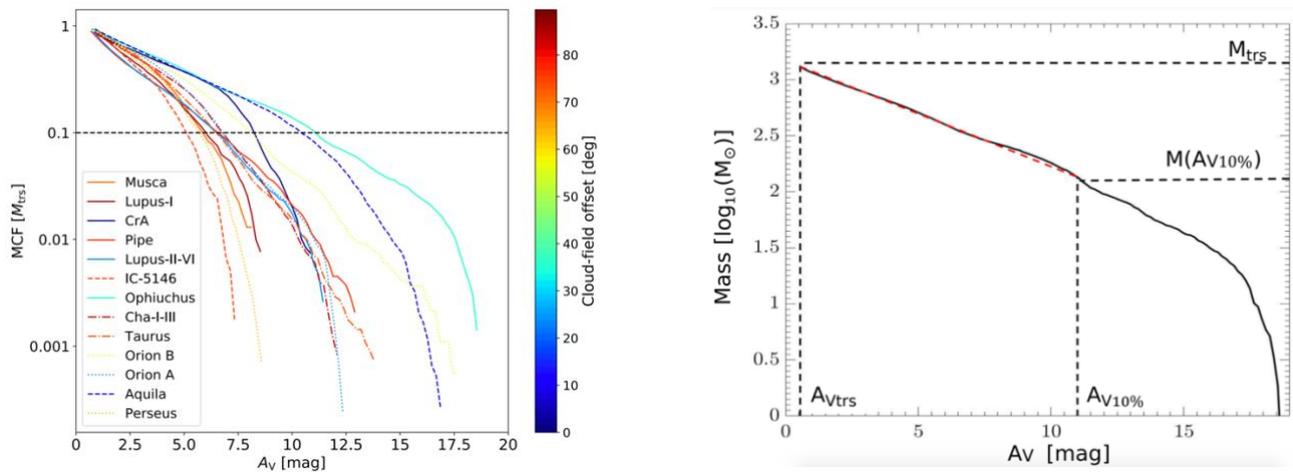

**Figure 8.** (**Left**): MCFs of all the Gould Belt clouds [34]. The MCFs are color coded by the cloud-field directional offsets. The starting point of each MCF is defined by the corresponding $A_{Vtrs}$ of the cloud. Perseus and Orion A/B are noted in dotted lines, because they have higher discrepancies of magnetic field directions inferred from optical and from PLANCK polarimetry data (Figure 6). Note that MCFs from smaller cloud-field offsets are in general shallower, which is further demonstrated in Figure 9. (**Right**): The definition of MCF slope illustrated with the case of Ophiuchus. The MCF slope (red dashed line) is defined by $\log_{10}(M_{trs})-\log_{10}(M_{Av10\%}) = 1$, divided by the difference between $A_{Vtrs}$ and $A_{V10\%}$.

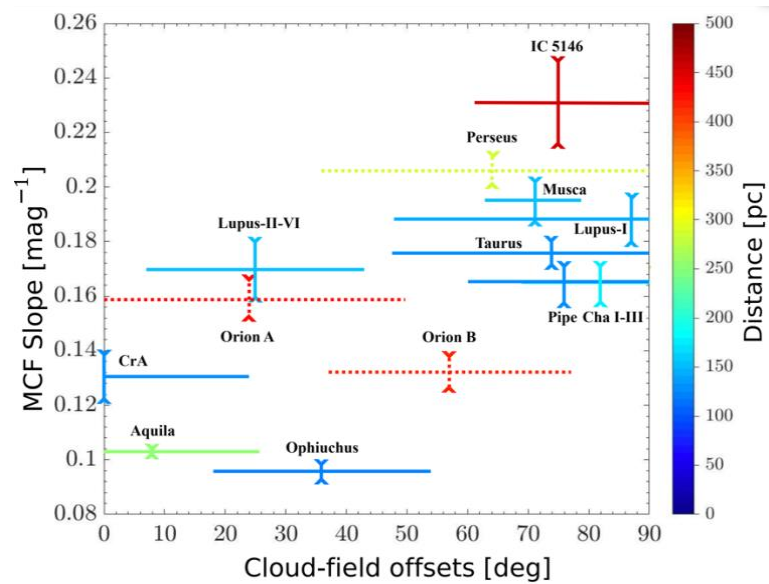

**Figure 9.** MCF slopes vs. cloud-field orientation offsets. Each symbol is color coded by the cloud distance (to show that resolution is not the cause of the trend). The cloud-field orientation offsets and the interquartile ranges (not the errors of the means) are adopted from [35]. The MCF uncertainties (vertical error bars) were defined in [34]. Perseus and Orion A/B (dotted lines) have more extensive line-of-sight scales and the largest discrepancies between the B-field directions inferred from Planck and optical data (Figure 6). The statistical tests summarized in Figure-7 caption were also performed here for a significant positive correlation [34].

## 4. Turbulence-Field Interaction

### 4.1. Turbulence Anisotropy

Given that cloud fields are ordered, turbulent velocities should be anisotropic. Turbulent energy cascades more easily in the direction perpendicular to the mean field than in that aligned with the field; see [36] for a short review.

Heyer et al. [37] first observed turbulence anisotropy in a ~2° × 2° diffuse region in the Taurus Molecular Cloud (TMC), where $A_v$ is less than 2 mag. Heyer and Brunt [4] extended the study of [37] to higher density using $^{13}CO$ (1–0) lines, which traces $A_v$ = 4–10 mag and exhibits little evidence of anisotropy. Heyer and Brunt [4] interpreted the observation as that the turbulence transits from sub- to super-Alfvénic in regions with higher column density. However, the regions observed by [4] still have very ordered B-field directions, unlike the interferometer-observed high-density regions discussed in Section 3. In Li et al. [36], we noted that $A_v$ = 4 mag is above the cloud-contraction threshold of TMC; see the grey range in Figure 5 (right), where the column density PDF transfers from log-normal to power-law-like. This means that, over this density limit, contraction velocities dominate the turbulent velocity, so turbulent anisotropy should be less observable.

**MHD Simulations**

Recently, Otto, Ji, and Li [38] performed the first numerical study of turbulence anisotropy involving gravity. They found that, from lines of sight with magnetic super-critical column densities, the degree of velocity anisotropy significantly dropped. Nevertheless, the Mach numbers of the super-/sub-critical regions are almost identical—both slightly increased from the initial condition—and stay sub-Alfvénic. The decrease in anisotropy with higher density, which is comparable to that observed by Heyer and Brunt [4], is due to the enhancement of the motion driven by gravity, not due to the increase in turbulent energy. Figure 10 [38] illustrates that the cores with highest aspect ratio, due to concentration along the B-field, have larger velocity dispersion along the field direction due to the flow concentration along the field. This contraction anisotropy is 90 degrees from turbulence anisotropy (which possesses the lowest velocity dispersion along the field). For a line-of-sight passing through both kinds of velocity anisotropy, it is more difficult to observe the anisotropy and, thus, the result is comparable to that of [4].

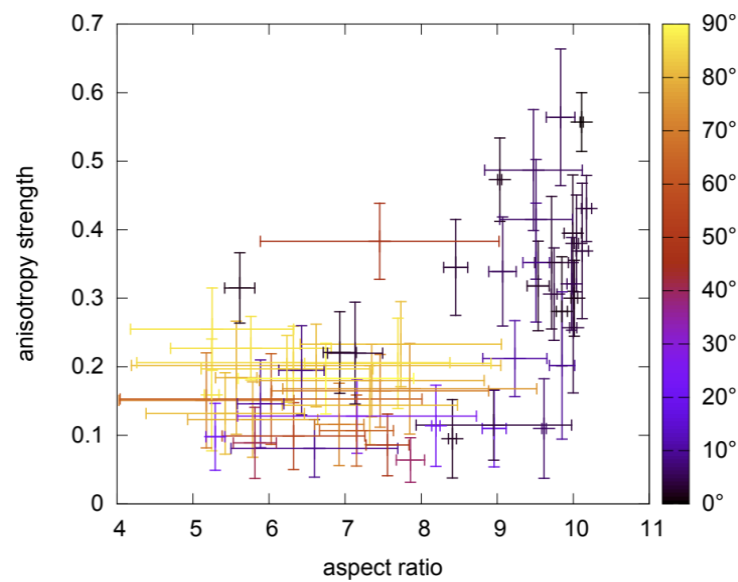

**Figure 10.** Correlation between velocity centroid anisotropy and column density autocorrelation structure. Each data point represents one snapshot from the gravitational stage of cloud simulations [38]. The horizontal axis shows the aspect ratio of the autocorrelation function from the top 10% contour of the column density. The vertical axis shows the degree of anisotropy (0–1) of the velocity centroid in the same region. The colour indicates the offset between the orientation of the column density autocorrelation and the orientation of the velocity centroid anisotropy (the orientation along which the centroid is most stable). Since the structure elongation is perpendicular to the B-field in the simulation, the velocity centroid anisotropy varies from being parallel to perpendicular with B-fields as density increases, while the dominant velocity changes from turbulence to contraction. This is consistent with the observation in Figure 11.

**Improved data analysis to reveal velocity anisotropy**

Turbulence anisotropy is a local behaviour [39]. The principal component analysis (PCA) carried out by [4] combined multiple velocity channels to form a set of orthogonal bases for the CO PPV (position–position–velocity) data cube. While the CO data possess a high spatial resolution in the P–P space, the combined velocity channels imply a large line-of-sight (LOS) scale as, for turbulence, $v \propto L^{\alpha}$, where $v$ is the velocity dispersion, which grows with channel numbers, and $L$ is the scale. This large LOS scale is disadvantageous for detecting turbulence anisotropy.

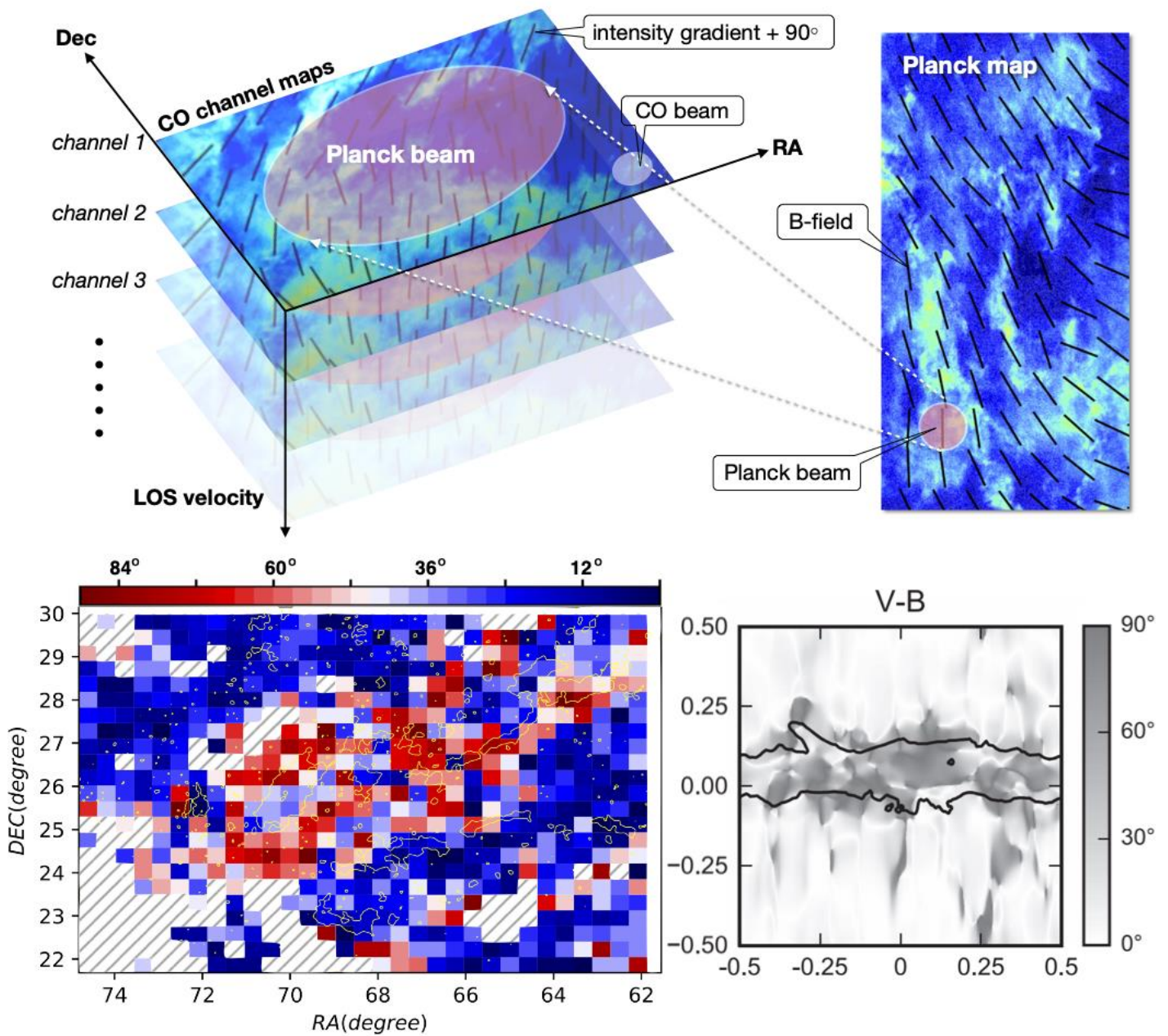

**Figure 11.** Velocity anisotropy in Taurus molecular cloud (TMC). (**Upper panel**) The method for detecting velocity anisotropy [40]. The intensity gradient of each CO velocity channel map is turned 90 degrees, in order to indicate the direction of velocity anisotropy. All the anisotropy directions covered by one Planck beam are averaged and weighted by intensity using the Stokes parameters (see Section 4.1), in order to compare with the B-field directions inferred from the Planck data. (**Lower-left**) The difference between B-field and velocity anisotropy directions. The yellow contours indicate the column density where the PDF inferred from the extinction data changes from log-normal to power-law [33]. The color code is for the directional difference between B-fields and velocity anisotropy. The red region (large directional difference) is coincident with the high-density region which is ~75° from the B-field (see Taurus in Figure 9). The striped areas are too

noisy to conclude. (**Lower-right**) Similar to the left panel but from one of the MHD simulations of [38]. They simulate slightly magnetic critical clouds with sub- to trans-Alfvénic turbulence. The contour shows the mean column density. The result is very similar to what was observed from Taurus—the directional offsets between velocity anisotropy and B-fields are small along LOS of low column densities but approaching perpendicularity as column density increases.

Based on the above argument, we tried a different way to detect velocity anisotropy [40]. We studied individual velocity channels to minimize the effective LOS scale (Figure 11). The anisotropy projected on each velocity channel resulted in elongated footprints; the elongation direction θ is perpendicular to the intensity gradient and, thus, can be derived from the channel maps. To average over all the velocity channels, we adopt the Stokes parameters used for averaging B-fields, as the goal is to compare velocity anisotropy with B-field directions. Q and U are defined in the same way as B-field analysis, only that the B-field directional angles are replaced by θ; that is, Q = cos(2θ) and U = sin(2θ). The resolution of the CO data is an order of magnitude higher than the B-field inferred from the Planck data, so all the Q's (or U's) within one Planck beam are averaged and weighted by the intensity to form q (or u). The mean direction can be derived from atan2(u/q)/2, while the square root of ($q^2 + u^2$) indicates the "degree of alignment" (DOA) of the θ's within the beam. The exemplary result from TMC is shown in Figure 11, where the offsets between B-field and velocity anisotropy vectors are shown. Small offsets between B-field and velocity anisotropy (blue in Figure 11) appear roughly in regions with lower column densities, while large offsets (red in Figure 11) are correlated with high column density. This is consistent with the MHD simulations (Figure 11; lower-right) [38], which showed that turbulence (dominant in low density) and contraction (dominant in high density) are both anisotropic due to the B-field, but in the perpendicular directions.

Through PCA, [4] concluded that velocity anisotropy only exists in the diffuse blue region within RA = 72–75 degrees in Figure 11. Our analysis reveals B-field related anisotropy almost everywhere. Note that the red region in Figure 11, in general, has higher DOA than blue regions. This means that the anisotropy is stronger here, compared to low-density regions; however, this is due to contraction, instead of turbulence.

*4.2. Turbulence-Induced Ambipolar Diffusion*

Much of the literature has claimed that ambipolar diffusion (AD)—the decoupling between the motions of neutrals and B-field—takes too long to be important in molecular clouds. Their arguments assumed $B \sim \mu G$. However, as can be seen in Figure 1, $B$ can be as high as mG in cloud cores, which greatly increases the ion-neutral drift velocity $v_d \propto B^2$ [13].

The field–turbulence interaction, discussed in Sections 3 and 4.1, ignores AD. In this section, however, I argue that AD should be expected in cloud cores and, indeed, it has been observed. The implication for star formation is significant, as turbulence is the only known angular momentum source for protostellar disks; however, the observed trans-Alfvénic turbulent energy is not capable of twisting B-fields to form disks if flux-freezing holds, which is called the "magnetic braking catastrophe".

**The Magnetic Reynolds number in molecular clouds**

The Magnetic Reynolds number, $R_M$ —the ratio of the advection term to diffusion term of the induction equation—is a dimensionless parameter that characterises how well ions and neutrals are coupled [41–46]. When $R_M \gg 1$, neutrals and ions are well-coupled. When $R_M$ is small, neutrals are no longer frozen into the B-fields and decouple from the charged species, which remain attached to the B-field lines due to the Lorentz force. A rough estimate of the decoupling scale, $l_{AD}$, can be calculated by setting $R_M = 1$. $R_M$ can be estimated by $L v /\eta$, where $\eta$ is the magnetic diffusivity and $v \propto L^\alpha$ for turbulence. Given any $\eta$, there always exists a $l_{AD}$, which is small enough to make $R_M$ unity, such that turbulent eddies smaller than $l_{AD}$ are not efficiently coupled to B-fields. Below the decoupling scale, ions and neutrals should have different turbulent velocity (energy) spectra, as illustrated in Figure 12A.

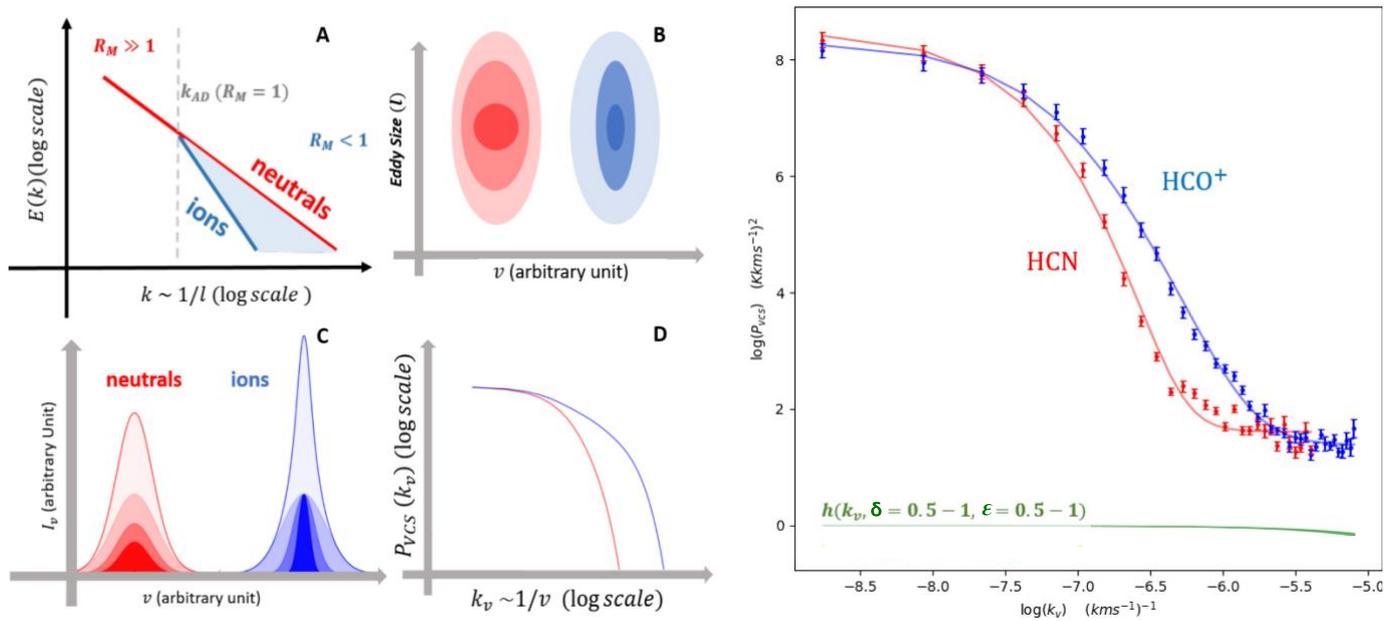

**Figure 12.** An illustration of turbulence energy spectra, emission line spectra and velocity coordinate spectra (VCS). *Panel A*: Illustration of a turbulence energy spectrum (in arbitrary units). The velocity dispersion/linewidth measured at scale *l* is proportional to the integration of the turbulent energies contributed by all eddies of size smaller than *l*. The turbulence energy in ions is dissipated steeply at scales smaller than $l_{AD}$, due to turbulence-induced AD. *Panel B*: Illustration of mapping turbulent eddies from real space to spectral lines (velocity space). In real space, three eddies with the same density have different sizes and velocity dispersion. The larger the velocity dispersion of an eddy, the larger the extent of it in the velocity space. Here, the largest eddy of these two species, neutrals (red) and ions (blue), share the same velocity dispersion. Due to the difference in the turbulent energy spectra (see *Panel A*), the energy carried by the small ions eddies decreases faster with the scale. *Panel C*: Illustration of Gaussian-type emission line spectra of the eddies in *Panel B*. The area under these Gaussians scales linearly with the sizes of the eddies in *Panel B*, while the velocity dispersions of the red and blue Gaussians are scaled with $\sigma_v \propto l^{1/3}$ and $\sigma_v \propto l$, respectively, to imitate the difference in the turbulence energy spectrum (*Panel A*). The eventual spectral lines (solid line) are given by the collective contributions of these profiles from individual eddies. *Panel D*: The VCSs of the spectral lines in *Panel C*. For ions, the spectral line in *Panel C* is narrower; thus, the VCS should contain more power in larger $k_v$. By studying the behaviour in Fourier space, VCS recovers the statistical description of the turbulence spectrum. *Right panel*: Comparable to *Panel D*, here the VCS obtained for HCN and HCO$^+$ (4-3) from NGC 6334 III in log-scale is shown. The best fits of the VCS are shown as solid lines. The effect of hyperfine structures enters as an extra factor in addition to the VCS; that is, $P_{hyper}(k_v) = h(k_v, \delta, \varepsilon) P_{VCS}(k_v)$ (see text). As it is presented in log–log space, the extra factor would simply up/downshift the VCS. The green line, $h(k_v, \delta, \varepsilon)$, demonstrates how the inclusion of the two brightest hyperfine structures would affect the behaviour of the VCS, with exaggerated intensities, $\delta$ and $\varepsilon$, ranging from 0.5 to 1 times of the main-line intensity. It is clear that the effect from $h$ is negligible.

For a weakly ionized plasma, such as a molecular cloud, the general form of Ohm's law is

$$E = -v_n \times B + (\eta j + \alpha j \times B + \beta j_\perp)$$

Where $v_n$ is the neutral velocity, $j$ is the electric current and, most importantly, $j_\perp$ is the current perpendicular to *B*. Furthermore,

$$\beta = \frac{B^2}{n_i \mu \nu_i}$$

is like the resistance due to the Lorentz force, where $n_i$ is the ion density, while $\nu_i$ and $\mu$ are the collision rate of an ion with the neutrals and the mean reduced mass characterizing such collisions, respectively. In the typical molecular cloud environment, $\beta \gg \alpha \gg \eta$, such that $\beta$ can be treated as the effective magnetic diffusivity [47]:

$$R_M \sim \frac{L v_n}{\beta}$$

with $B \sim 30$ μG, $n_n \sim 10^3$ cm$^{-3}$, ionization fraction $\sim 10^{-6}$, and $v_i \sim 1.5 \times 10^{-9} n_n$ s$^{-1}$ [48], $\beta \sim 10^{22}$ cm$^2$/s for a typical molecular cloud. On the other hand, $v_n \sim (L/1\text{pc})^{0.5}$ km/s (e.g., [49,50]), so, when $L = 0.1$ pc, $R_M \sim 1$; that is, $l_{AD} \sim 0.1$ pc, which is the typical size of a cloud core.

**First observation**

The spectrum difference shown in Figure 12A should imprint into the spectral line-width (Figure 12B), and we should expect different line-widths for ions and neutrals. As all the turbulent eddies within a telescope beam contribute to the observed line-width (velocity dispersion), the decoupling should affect the line-width, even if the beam size is larger than $l_{AD}$. Indeed, many other factors (e.g., opacities, spatial distributions, hyperfine structures, and outflows) can affect line-widths; however, these non-magnetic factors can be ruled out by carefully choosing the regions and tracers of observations. HCO$^+$/HCN can be used for studying AD [41], as the ion species generally have a higher opacity, slightly more extended distribution, and more unresolved hyperfine structures (for the J = 1–0 transition cases [51]). All of these factors tend to widen the line-width, yet ion spectra are systematically narrower [52–54]. Most importantly, we [51] have shown that the line-width difference between HCO$^+$ and HCN is proportional to the B-field strength. This definitely cannot be explained by any reason other than the decoupling between B-fields and small turbulent eddies. This decoupling has been called turbulence-driven ambipolar diffusion (TAD) [41] in order to distinguish it from gravity-driven ambipolar diffusion—the decoupling between B-fields and gravitationally contracting gas [20,55].

**First direct evidence of $l_{AD}$**

While the B-field dependent line-width difference between ions and neutrals can only be explained by TAD, the $l_{AD}$ cannot be derived simply from line-width differences. To directly identify $l_{AD}$, we have to find a scale below which ions and neutrals possess different turbulent energy spectra, as shown in Figure 12A. To identify turbulent spectra within a molecular cloud is not an easy task, as not only the velocity field, but also the density profile can contribute to the observed line-width. One promising method to disentangle the contributions of velocity and density is the method called the velocity co-ordinate spectrum (VCS). A detailed derivation of the VCS can be found in [56]. The spirit of the method is very similar to how astronomers study the cosmic H I spatial distribution. The density-scale relation is obtained from studying the 21 cm intensity along the frequency domain, based on a known connection between the scale and frequency—Hubble's law. Similarly, the known Kolmogorov-type scale–velocity relation of turbulence has an imprint on the emission line intensity along the velocity domain.

The VCS is defined as the power spectrum of an emission line in the velocity coordinate:

$$P_{VCS}(k_v) = \left| \int_{-\infty}^{\infty} I(v) e^{i k_v v} dv \right|^2$$

, where $I(v)$ is the spectral line intensity, and $k_v$ is the wavenumber in the velocity coordinate ($k_v = 2\pi/v$). Examples of $P_{VCS}$ are shown in Figure 12. It can be shown that $P_{VCS}$ is, not surprisingly, a function of the turbulent energy spectrum, $E(k)$, the column density spectrum $N(k)$, the temperature ($T$), and the line-of-sight scale of the cloud ($L_C$), where $k$ is the spatial wave number. The fact that HCO$^+$ and HCN are highly spatially correlated implies that $T$ and $L_C$ are close to identical for these two species. It has been observed that $E(k) \propto k^{-\alpha}$ (Kolmogorov-type) and $N(k) \propto k^{-\gamma}$, with $\gamma = 4$–5 for typical molecular clouds (e.g., [57]). A critical point that [56] concluded upon is that, when $\gamma > 3$, the effect of $N(k)$ on $P_{VCS}$ is negligible, compared to that of $E(k)$. Therefore, the apparent difference between the two $P_{VCS}$ in Figure 12 must be due to the difference in $\alpha$. With $P_{VCS}$(HCN) and $T$ as the observables, we can fit the $P_{VCS}$ to obtain $\alpha$(HCN) and $L_C$. Then, with $P_{VCS}$(HCO$^+$), $T$, $L_C$, and assuming $\alpha$(HCO$^+$) = $\alpha$(HCN) above $l_{AD}$, we can fit $P_{VCS}$ to obtain $l_{AD}$ and $\alpha$(HCO$^+$) below this scale. For NGC 6334 III (see Figure 12), we found that $l_{AD} \simeq 0.48 \pm 0.16$ pc, $\alpha$(HCN) $\simeq 1.66 \pm 0.09$, and $\alpha$(HCO$^+$) below $l_{AD} \simeq 2.01 \pm 0.05$ [46].

**The effect of opacity and hyperfine structures**

We can estimate optical depths with RADEX [58]. The HCO+ (3–2) data of NGC 6334 has been best-fitted to the model of a cloud with a fractional abundance of HCO+, $X(HCO^+)$ = 2 × 10$^{-9}$ [59]. Assuming a column density of $N(H) \approx 8.5 \times 10^{22}$ cm$^{-2}$ [60], this gives a column density of $N(HCO^+) \approx 1.7 \times 10^{14}$ cm$^{-2}$. For the (4–3) transition, HCO+ is marginally optically thin, with an optical depth ($\tau$) of 0.86. Although there is no existing measurement of HCN abundance in this particular cloud, we can estimate $N(HCN)$ with the abundance ratio between HCN and HCO+ found in general star-forming regions: $N(HCN)/N(HCO^+) \approx 2$ [61]. This gives $\tau(HCN) = 0.127$, which is very optically thin. This means that HCN is better at tracing the entire range of the turbulent spectrum and can offer a good constraint on the input of $L_C$ and large-scale turbulent spectrum for the modelling of HCO+. As long as the velocity scale at $l_{AD}$ is smaller than $v(L_C)/\tau$ (which is true here, as $\tau < 1$), which is the upper-limit for VCS not affected by $\tau$, the result for HCO+ is also valid [62].

The HCN transitions from rotational level J = 4–3 has five hyperfine transitions [63]. Assuming local thermal equilibrium, the two brightest satellite lines are 0.12 km/s and 0.038 km/s, respectively, from the main transition. This offset cannot account for our observed line-width difference between HCO+ and HCN (~0.4 km/s). The effect of hyperfine structures on the application of VCS has also been studied [46]. We assumed the main line as $I(v)$. Then, the spectral line with the two brightest satellite lines can be described by $I_{hyper}(v) = \delta I(v-a) + I(v) + \varepsilon I(v+b)$, where $\delta$ and $\varepsilon$ are the relative peak intensities of the two satellite lines, compared to the main line; $a$ and $b$ are the shifts of the hyperfine satellites on the velocity axis, with respect to the main line. The resulting power spectrum of the spectral lines in the presence of hyperfines is in the form of

$$\begin{aligned} P_{hyper}(k_v) &= \left| \int_{-\infty}^{\infty} I_{hyper}(v) e^{ik_v v} dv \right|^2 \\ &= [1 + \delta^2 + \varepsilon^2 + 2\delta \cos(k_v a) + 2\varepsilon \cos(k_v b) \\ &\quad + 2\delta\varepsilon \cos(k_v(a+b))] P_{VCS}(k_v) \\ &= h(k_v, \delta, \varepsilon) P_{VCS}(k_v) \end{aligned}$$

The resulting power spectrum has an additional factor, $h(k_v, \delta, \varepsilon)$, which is demonstrated in Figure 12. Note that the inclusion of $h$ simply shifts $P_{VCS}(k_v)$ and does not change the shape of the VCS if $h$ is rather uniform, which is the case shown in Figure 12.

**5. Summary**

Star formation is a result of the competition between gravity, B-fields and turbulence affecting the gas distribution in molecular clouds. Due to the diffusion and straightening nature of B-field lines, which act against gravity's concentration and turbulence's randomization effects on gas, the compressibility ($B$–$\rho$ relation) and orderliness of B-field lines are considered as indications of the B-field's dynamical importance, relative to gravity and turbulence.

In turn, gravitational contraction and turbulence velocity must be anisotropic, if B-fields are dynamically important.

**B-field vs. gravity**

The temporal $B$–$\rho$ relation from the textbook model (Figure 1, upper-left) is not observable. The observed spatial $B$–$\rho$ relation inferred from Bayesian analysis on the Zeeman measurements is not only incomparable to the textbook model, but also inconclusive, due to the uncertainties of the existing detections of $B$ and $\rho$ (i.e., errors-in-variables bias).

On the other hand, the $B$–$N$ relation (Figure 1, lower-right) indicates that cloud-core column densities stay very close to the magnetic critical values and the lower densities ($N < 10^{22}$/cm$^2$) are magnetic sub-critical. This fact is in agreement with the observations that cloud shapes (Figure 5), fragmentation (Figure 9), and star formation efficiencies (Figure 7) are all regulated by field orientations.

**B-field vs. turbulence**

By comparing the B-field orientation dispersions from observations and from MHD simulations at various scales, we realized that molecular cloud turbulence is sub-Alfvénic at cloud scales and turns slightly super-Alfvénic within sub-pc cloud cores (Figures 3 and 4). Sub-Alfvénic turbulence must be anisotropic, so should be magnetically trans-critical contraction, as illustrated by MHD simulations (Figure 10) and TMC observations (Figure 11). However, the two types of anisotropy dominate at different densities and are orthogonal to each other.

The slightly super-Alfvénic turbulence in cloud cores is not enough to twist the B-field and form protostellar disks; the so-called magnetic braking catastrophe. This may, however, be an overconcern, as flux-freezing is not a good approximation in cloud cores where the magnetic Reynolds number is not very high. Turbulence-induced ambipolar diffusion (TAD)—the decoupling between neutral eddies and B-field—has recently been observed below 0.4 pc (Figure 12), which indicates that ideal MHD is not ideal for cloud core simulations.

**Funding:** The research summarized here was supported by the Research Grants Council of Hong Kong: General Research Funds 24300314, 14600915, 14307118, 14307019, 14305717 and 14304616; Collaborative Research Fund C4012-20E.

**Acknowledgments** The author is grateful for the students and teachers (Drs. Po Kin Leung, Frank Otto and Hsiang Hsu Wang) of the CUHK Star Formation Group and Prof. Xiaodan Fan of CUHK Dept. of Statistics. The article summarizes part of their collaborative efforts from 2014 to 2020.

**Conflicts of Interest:** The author declares no conflict of interest.